\documentclass{opt2022} 

\usepackage{lastpage}
\usepackage{multirow}
\usepackage{booktabs}
\usepackage{setspace}
\usepackage{relsize}
\usepackage{physics}
\usepackage{float}
\usepackage{footnote}
\usepackage{amssymb}
\usepackage{pifont}
\usepackage{xcolor}
\usepackage{amsmath}
\usepackage{amssymb}
\usepackage{mathtools}
\usepackage{enumitem}
\usepackage[textsize=tiny]{todonotes}
\usepackage[noamssymbols,upint]{newtxmath} 

\usepackage{comment}
\hypersetup{
    colorlinks=true,
    linkcolor=red,
    citecolor=blue,
    filecolor=magenta,      
    urlcolor=blue,
linktocpage}
\newcommand{\specialcell}[2][c]{%
\begin{tabular}[#1]{@{}c@{}}#2\end{tabular}}
\newcommand{\cmark}{\textcolor{green}{\ding{51}}}%
\newcommand{\xmark}{\textcolor{red}{\ding{55}}}%
\def\semichecked{\textcolor{blue}{\checkmark\!\!\!\!\raisebox{0.3 em}{\tiny$\smallsetminus$}}}
\newcommand{\halfcmark}{\textcolor{green}{\ding{51}}\textsuperscript{\textcolor{red}{\kern-0.5em\tiny\ding{55}}}}

\usepackage[frozencache,cachedir=.]{minted}
\newminted{python}{frame=lines,framerule=2pt}

\newcommand{\torchopt}{\texttt{TorchOpt}}

\title[Ren, Feng, Liu, Pan, Fu, Mai, and Yang]{\texttt{TorchOpt}: An Efficient Library for Differentiable Optimization}

\optauthor{\Name{Jie Ren}$^{1,}$\footnotemark[1] \Email{jieren9806@gmail.com}\\
\Name{Xidong Feng}$^{2,}$\footnotemark[1] \Email{xidong.feng.20@ucl.ac.uk}   \\
\Name{Bo Liu}$^{3,}$\footnotemark[1] \Email{benjaminliu.eecs@gmail.com}\\ 
\Name{Xuehai Pan}$^{4,}$\footnotemark[1] \Email{xuehaipan@pku.edu.cn}\\
\Name{Yao Fu}$^{1}$ \Email{y.fu@ed.ac.uk}\\ 
\Name{Luo Mai}$^{1,}$\footnotemark[2] \Email{luo.mai@ed.ac.uk}\\
\Name{Yaodong Yang}$^{4,}$\footnotemark[2]  \Email{yaodong.yang@pku.edu.cn} \\
\addr $^1$ University of Edinburgh, Edinburgh, United Kingdom \\
\addr $^2$ University College London, London, United Kingdom \\
\addr $^3$ Institute of Automation, Chinese
Academy of Sciences, Beijing, China \\
\addr $^4$ Institute for AI, Peking University, Beijing, China
}

\begin{document}

\maketitle

\begin{abstract}
Recent years have witnessed the booming of various differentiable optimization algorithms.
These algorithms exhibit different execution patterns, and their execution needs massive computational resources that go beyond a single CPU and GPU.
Existing differentiable optimization libraries, however, cannot support efficient algorithm development and multi-CPU/GPU execution, making the development of differentiable optimization algorithms often cumbersome and expensive. This paper introduces \texttt{TorchOpt}, a PyTorch-based efficient library for differentiable optimization. \texttt{TorchOpt} provides a unified and expressive differentiable optimization programming abstraction. This abstraction allows users to efficiently declare and analyze various differentiable optimization programs with explicit gradients, implicit gradients, and zero-order gradients.
\texttt{TorchOpt} further provides a high-performance distributed execution runtime. This runtime can fully parallelize computation-intensive differentiation operations (e.g. tensor tree flattening) on CPUs / GPUs and automatically distribute computation to distributed devices. Experimental results show that \texttt{TorchOpt} achieves $5.2\times$ training time speedup on an 8-GPU server. \texttt{TorchOpt} is available at: \url{https://github.com/metaopt/torchopt}.

\renewcommand{\thefootnote}{\fnsymbol{footnote}}
\footnotetext[1]{Equal contribution.}
\footnotetext[2]{Corresponding author.}
\renewcommand*{\thefootnote}{\arabic{footnote}}
\renewcommand{\thefootnote}{\fnsymbol{footnote}}
\end{abstract}

\section{Introduction}

Recent years have witnessed the booming of differentiable optimization-based algorithms, including MAML~\cite{finn2017model}, OptNet~\cite{amos2017optnet}, and MGRL~\cite{xu2018meta}. One of the important parts of differentiable optimization is meta-gradient, which is the gradient term of outer-loop variables by differentiating through the inner-loop optimization process. By leveraging meta-gradients, machine learning models can increase the sample efficiency~\cite{finn2017model} and the final performance~\cite{xu2018meta}.

Developing differentiable optimization algorithms poses several challenges. First, developers need to realize different inner-loop optimization and implement algorithms with gradient flows on complex computational graphs. Examples include explicit gradient computation of
unrolled optimization~\cite{finn2017model,xu2018meta}, implicit gradient for differentiable optimization~\cite{amos2017optnet,cvxpy}, evolutionary strategies for non-differentiable optimization~\cite{feng2021neural}, adjoint methods for differentiable ordinary differentiable equations~\cite{chen2018neural}, Gumbel-Softmax to differentiate through discrete distribution~\cite{jang2016categorical}, and function interpolation for differentiable combinatorial solvers~\cite{poganvcic2019differentiation}, etc. 
Second, differentiable optimization is computation-intensive. The meta-gradient computation requires heavy Hessian computation~\cite{finn2017model}, high-dimensional linear equations~\cite{imaml}, or large task-level batch size~\cite{oh2020discovering}. Such computation requirement often goes beyond what a single CPU and GPU can provide. 

None of the existing differentiable optimization libraries can provide
full support for efficient algorithm development and execution.
Most libraries target a limited number of differentiable optimizers~\cite{higher,torchmeta,optax,learn2learn}. They cannot fully support implicit differentiation~\cite{jaxopt,hypertorch,betty}, zero-order gradient~\cite{betty}, and distributed training~\cite{learn2learn,jaxopt}, as shown in Table~\ref{tab:compare}. As a result, researchers have to implement algorithms in an ad-hoc and application-specific manner, making the development process cumbersome and expensive. Further, critical system optimization techniques (e.g. GPU optimization and distributed execution) are tightly coupled with certain algorithms and they are hard to be enabled for all possible algorithms.

To address these issues, this paper introduces \texttt{TorchOpt}, a PyTorch library that makes it efficient to develop and execute differentiable optimization algorithms with multiple GPUs. 
The design and implementation of \texttt{TorchOpt} make the following contributions:

\paragraph{(1) Unified and expressive differentiation mode for differentiable optimization.} \texttt{TorchOpt} provides a general set of low-level / high-level / functional / Object-Oriented (OO) API to help users flexibly enable differentiable optimization within the computational graphs produced by PyTorch. Specifically, \texttt{TorchOpt} supports three differentiation modes for handling differentiable optimization problems: (i)~Explicit gradient for unrolled optimization, (ii)~implicit gradient for differentiable optimization, and (iii)~zero-order gradient estimation for non-smooth/differentiable functions.

\paragraph{(2) High-performance and distributed execution runtime.} \texttt{TorchOpt} aims to enable differentiable optimization algorithms to fully utilize CPUs and GPUs. To achieve this, we design (i) CPU/GPU accelerated optimizers (e.g., SGD, RMSProp, Adam) that realize the fusion of small differentiable operators and a full offloading of these operators to GPUs, (ii) parallel OpTree which can fully parallelize the nested structure flattening (Tree Operations), a key computation-intensive operation in differentiable optimization, on distributed CPUs, and (iii) a distributed auto-grad framework which can automatically identify the inner-loop tasks in differentiable optimizers and dispatch the execution of inner-loop tasks to distributed CPUs and GPUs.

Experimental results show that \texttt{TorchOpt} can reduce PyTorch optimizer forward/backward time, by $5\times$ to $10\times$ on CPU and $5\times$ to $20\times$ on GPU. \texttt{TorchOpt} can reduce the training time of the MAML~\cite{finn2017model} algorithm by $5.2\times$ by distributing MAML computation to 8 GPUs.

\begin{table}[htbp]
  \begin{center}
    \begin{scriptsize}
      \caption{Comparison of \texttt{TorchOpt} and other differentiable optimization libraries. Note: $\semichecked$ indicates
        that the feature is partially supported.}
      \begin{tabular}{l c c c c c c c}
        \toprule
                                                & \specialcell{Differentiable                                                            \\Optimizer} & \specialcell{Implicit \\ Differentiation} & \specialcell{Zero-order\\Gradient} & \specialcell{Accelerated\\Operator} & \specialcell{Distributed\\Training} & \specialcell{Debugging\\Support} & \specialcell{Backend}\\
        \midrule
        \texttt{higher}~\cite{higher}           & \cmark                      & \xmark       & \xmark & \xmark & \xmark       & \xmark &
        PyTorch                                                                                                                          \\
        \texttt{Optax}~\cite{optax}             & \cmark                      & \xmark       & \xmark & \xmark & \xmark       & \xmark &
        JAX                                                                                                                              \\
        \texttt{Torchmeta}~\cite{torchmeta}     & \cmark                      & \xmark       & \xmark & \xmark & \xmark       & \xmark &
        PyTorch                                                                                                                          \\
        \texttt{learn2learn}~\cite{learn2learn} & \cmark                      & \xmark       & \xmark & \xmark & \semichecked & \xmark &
        PyTorch                                                                                                                          \\
        \texttt{JAXopt}~\cite{jaxopt}           & \cmark                      & \cmark       & \xmark & \xmark & \cmark       & \xmark &
        JAX                                                                                                                              \\
        \texttt{HyperTorch}~\cite{hypertorch}   & \cmark                      & \cmark       & \xmark & \xmark & \xmark       & \xmark &
        PyTorch                                                                                                                          \\
        \texttt{Betty}~\cite{betty}             & \cmark                      & \semichecked & \cmark & \xmark & \semichecked & \xmark &
        PyTorch                                                                                                                          \\
        \texttt{TorchOpt} (ours)                & \cmark                      & \cmark       & \cmark & \cmark & \cmark       & \cmark &
        PyTorch                                                                                                                          \\
        \bottomrule
      \end{tabular}
    \end{scriptsize}
  \end{center}
  \label{tab:compare}
  \vspace{-1em}
\end{table}

\raggedbottom

\section{\texttt{TorchOpt} Design and Implementation}

\subsection{Architecture Overview}

\begin{figure}[t]
    \centering
    \includegraphics[width=0.95\linewidth]{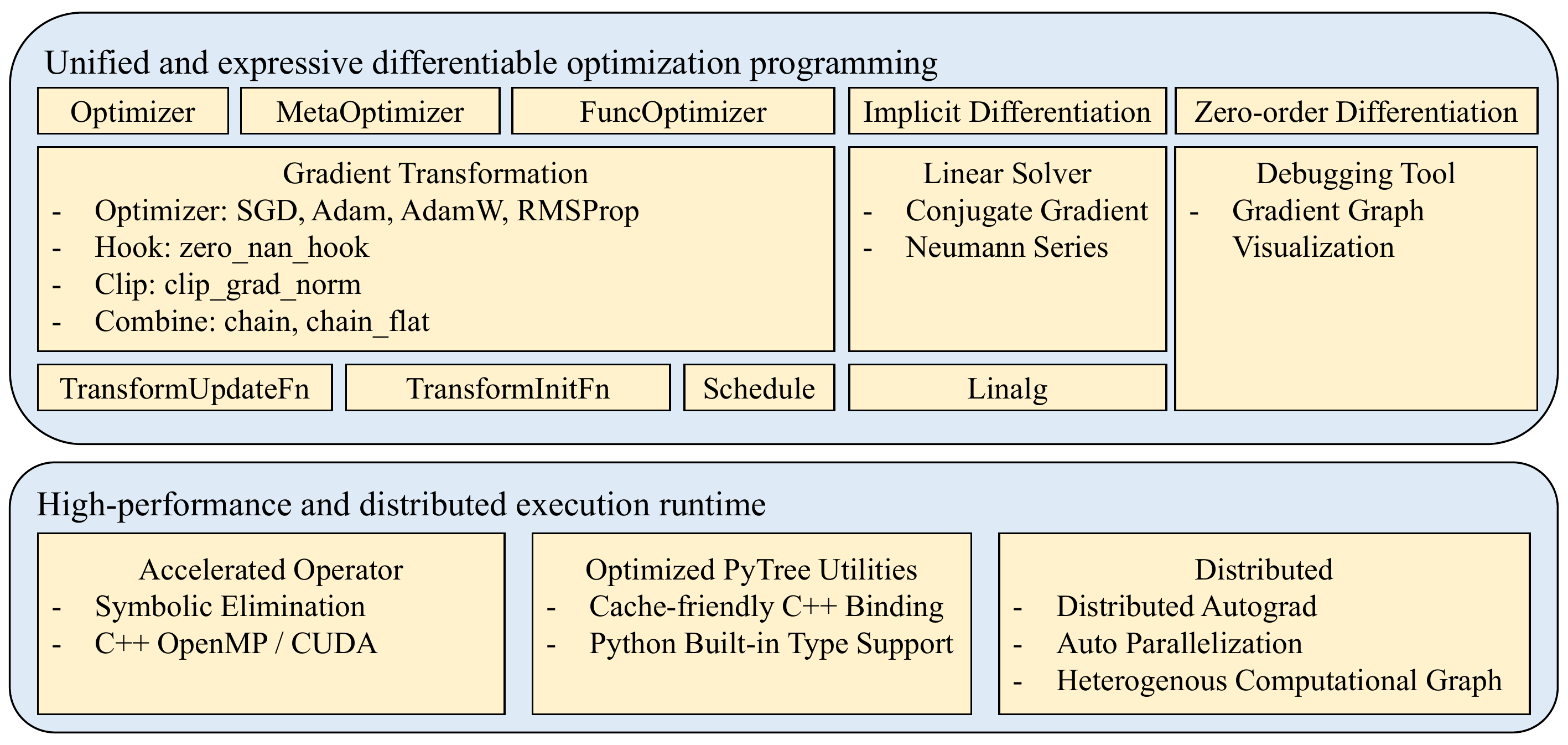}
    \caption{\texttt{TorchOpt}'s architecture overview. 
    }
    \label{fig:overview}
\end{figure}

Figure~\ref{fig:overview} gives an overview of the system architecture, \torchopt~consists of two different aspects, the unified and expressive differentiable optimization programming lets users easily implement differentiable optimization algorithms, we provide both high-level APIs and low-level APIs for three differentiation modes along with debugging tools, all of which are described in Sec.~\ref{unifiedoptimization}. Then the high-performance and distributed execution runtime contains several accelerated solutions to support fast differentiation with different modes on GPU \& CPU and distributed training features for multi-node multi-GPU scenario, which we demonstrate boost performance in Sec.~\ref{high-performance}. Additionally, we offer \texttt{OpTree} to enable fast structure \texttt{flatten} and \texttt{unflatten}, which is specially designed for our functional programming implementation. We use an optimized structure to avoid memory allocation if the sub-tree is small.
\begin{figure}[t]
    \centering
    \includegraphics[width=0.9\linewidth]{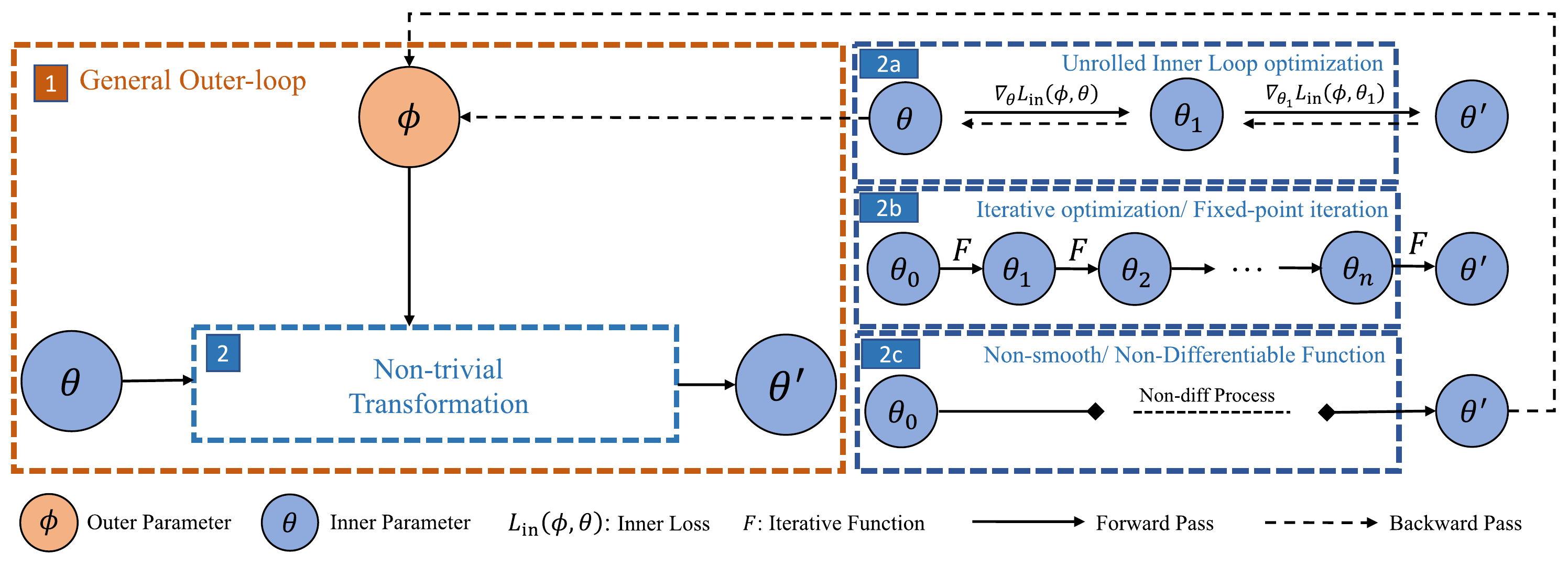}
    \caption{\texttt{TorchOpt}'s differentiation modes. By formulating the problem as a differentiable problem, \texttt{TorchOpt} offers Autograd support for the backward pass (dotted lines).}
    \label{fig:diffmode}
    \vspace{-2em}
\end{figure}

\subsection{Programming Abstraction}
\label{unifiedoptimization}
\texttt{TorchOpt} aims to provide (i)~high-level APIs that allow users to directly import differentiable optimizers, (ii)~low-level APIs that enable automatic differentiation in different applications, and (iii)~tools that allow users to analyze gradient flow in gigantic computational graphs.

The key challenge of consolidating these high-level and low-level APIs in a single library is that we must have a unified abstraction that allows different differentiable optimization algorithms to be easily declared. To address this, we design a differentiable optimization updating scheme, which can be easily extended to realize various differentiable optimization processes. 

As shown in Fig.~\ref{fig:diffmode}, the scheme contains an outer level that has parameters $\boldsymbol{\phi}$ that can be learned end-to-end through the inner level parameters solution $\boldsymbol{\theta}^{\prime}(\boldsymbol{\phi})$ (treating solution $\boldsymbol{\theta}^{\prime}$ as a function of $\boldsymbol{\phi}$) by using the best-response derivatives $\partial \boldsymbol{\theta}^{\prime}(\boldsymbol{\phi})/ \partial \boldsymbol{\phi}$. It can be seen that the key component of this algorithm is to calculate the best-response (BR) Jacobian. From the BR-based perspective, \texttt{TorchOpt} supports three differentiation modes: explicit gradient over unrolled optimization, implicit differentiation, and zero-order differentiation.

\noindent \textbf{Explicit Gradient (EG) over unrolled optimization.} As shown in Fig.~\ref{fig:diffmode}-2a, the idea of EG is to treat the gradient step as a differentiable function and try to backpropagate through the unrolled optimization path. This differentiation mode is suitable for algorithms when the inner-level optimization solution is obtained by a few gradient steps, such as MAML~\cite{finn2017model}, MGRL~\cite{xu2018meta}. \texttt{TorchOpt} offers both functional and OOP API. Refer to Listing \ref{eg} for the code snippet.

\begin{listing}[!ht]
\begin{minipage}{0.59\textwidth}
\begin{minted}[frame=single,framesep=2pt, fontsize={\relscale{0.625}}]{python}
# Functional API
opt = torchopt.adam()
# Define meta and inner parameters
meta_params = ...
fmodel, params = make_functional(model)
# Initialize optimizer state
state = opt.init(params)

for iter in range(iter_times):
    loss = inner_loss(fmodel, params, meta_params)                
    grads = torch.autograd.grad(loss, params)
    # Apply non-inplace parameter update
    updates, state = opt.update(grads, state, inplace=False)
    params = torchopt.apply_updates(params, updates)   

loss = outer_loss(fmodel, params, meta_params)
meta_grads = torch.autograd.grad(loss, meta_params)
\end{minted}
\end{minipage}
\begin{minipage}{0.41\textwidth}
\begin{minted}[frame=single,framesep=2pt, fontsize={\relscale{0.625}}]{python}
# OOP API
# Define meta and inner parameters
meta_params = ...
model = ...
# Define differentiable optimizer
opt = torchopt.MetaAdam(model)

for iter in range(iter_times):
    # Perform the inner update
    loss = inner_loss(model, meta_params)  
    opt.step(loss)

loss = outer_loss(model, meta_params)
loss.backward()



\end{minted}
\end{minipage}
\caption{\texttt{TorchOpt} code snippet for explicit gradient.}
\label{eg}
\end{listing}

\noindent \textbf{Implicit Gradient (IG).} As shown in Fig.~\ref{fig:diffmode}-2b, by treating the solution $\boldsymbol{\theta}^{\prime}$ as an implicit function of $\boldsymbol{\phi}$, the idea of IG is to directly get analytical best-response derivatives $\partial \boldsymbol{\theta}^{\prime}(\boldsymbol{\phi})/ \partial \boldsymbol{\phi}$ by implicit function theorem~\cite{krantz2002implicit}. This is suitable for algorithms when the inner-level optimal solution is achieved (${\left. \frac{\partial F(\boldsymbol{\theta},\boldsymbol{\phi})}{\partial \boldsymbol{\theta}} \right\rvert}_{\boldsymbol{\theta}^{\prime}} = 0$) or reaches some stationary conditions ($F(\boldsymbol{\theta}^{\prime},\boldsymbol{\phi})=0$), such as iMAML~\cite{imaml}, DEQ~\cite{bai2019deep}. \texttt{TorchOpt} offers functional/OOP API for supporting both conjugate gradient-based~\cite{imaml} and Neumann series~\cite{lorraine2020optimizing} based method. Refer to Listing \ref{ig} for the code snippet.

\begin{listing}[!ht]
\begin{minipage}{0.49\textwidth}
\begin{minted}[frame=single,framesep=2pt, fontsize={\relscale{0.6}}]{python}
# Functional API for implicit gradient
def stationary(params, meta_params, batch, labels):
    # Stationary condition construction
    ...
    return stationary condition

@torchopt.diff.implicit.custom_root(stationary)
def solve(params, meta_params, batch, labels):
    # Forward optimization process
    ...
    return optimal_params



\end{minted}
\end{minipage}
\begin{minipage}{0.51\textwidth}
\begin{minted}[frame=single,framesep=2pt, fontsize={\relscale{0.6}}]{python}
# OOP API
class Module(torchopt.nn.ImplicitMetaGradientModule):
    def __init__(self, meta_module, ...):
      ...
    def forward(self, x):
      # Forward process
      ...
    def optimality(self, batch, labels):
      # Stationary condition construction
      ...
    def solve(self, batch, labels):
      # Forward optimization process
      ...
      return self
\end{minted}
\end{minipage}
\caption{\texttt{TorchOpt} code snippet for implicit gradient.}
\label{ig}
\end{listing}

\noindent \textbf{Zero-order Differentiation (ZD).} As shown in Fig.~\ref{fig:diffmode}-2c, when the inner-loop process is non-differentiable or one wants to eliminate the heavy computation burdens in the previous two modes (brought by Hessian), one can choose ZD. ZD typically gets gradients based on zero-order estimation, such as finite-difference, or Evolutionary Strategy (ES)~\cite{salimans2017evolution}. ESMAML~\cite{song2019maml}, and NAC~\cite{feng2021neural}, successfully solve the differentiable optimization problem based on ES. Instead of optimizing the objective $F$, ES optimizes a Gaussion smoothing objective defined as $\tilde{f}_\sigma(\boldsymbol{\theta}) = \mathbb{E}_{\vb*{z} \sim \mathcal{N} \qty( \vb*{0}, \vb*{I}_d )} \qty[ f (\boldsymbol{\theta} + \sigma \vb*{z}) ]$, where $\sigma$ denotes precision. The gradient of such objective is $\nabla_\theta \tilde{f}_\sigma(\theta) = \frac{1}{\sigma} \mathbb{E}_{\vb*{z} \sim \mathcal{N} \qty( \vb*{0}, \vb*{I}_d )} \qty[ f(\boldsymbol{\theta} + \sigma \vb*{z}) \vb*{z} ]$. \texttt{TorchOpt} also offers functional and OOP API for ES method. Refer to Listing \ref{es} for code snippets.

\begin{listing}[!ht]
\begin{minipage}{0.495\textwidth}
\begin{minted}[frame=single,framesep=2pt, fontsize={\relscale{0.6}}]{python}
# Functional API
# Customize the noise sampling function in ES
def sample(params, batch, labels, *, sample_shape):
    ...
    return sample_noise
# Specify the method and parameter of ES
@torchopt.diff.zero_order(method, sample)
def forward(params, batch, labels):
    # Forward process
    return output
\end{minted}
\end{minipage}
\begin{minipage}{0.505\textwidth}
\begin{minted}[frame=single,framesep=2pt, fontsize={\relscale{0.6}}]{python}
# OOP API
class ESModule(torchopt.nn.ZeroOrderGradientModule):
    def sample(self, batch, labels, sample_shape):
      # Customize the noise sampling function in ES
      ...
      return sample_noise
    def forward(self, batch, labels):
      # Forward process
      ...
      return output
\end{minted}
\end{minipage}
\caption{\texttt{TorchOpt} code snippet for zero-order differentiation.}
\label{es}
\end{listing}

\vspace{-0.5em}

\noindent \textbf{Gradient graph visualization}. Complex gradient graph in meta-learning/differentiable optimization brings a great challenge for managing the gradient graph and debugging the code. \texttt{TorchOpt} provides a visualization tool that draws variable (e.g. network parameters or meta parameters) names on the gradient graph for better analysis. The visualization tool is modified from TorchViz~\cite{torchviz}. Compared with TorchViz, \texttt{TorchOpt} fuses the operations within the optimization algorithm (such as Adam) to reduce the complexity and provide simpler visualization. Refer to the visualization example in Appendix \ref{apx:visualization}. 

\subsection{High-performance and Distributed Runtime}\label{high-performance}

\begin{figure}[tb]
    \centering
    \vspace{-0.5em}
    \subfigure[CPU-accelerated optimizer]{\includegraphics[width=.31\textwidth]{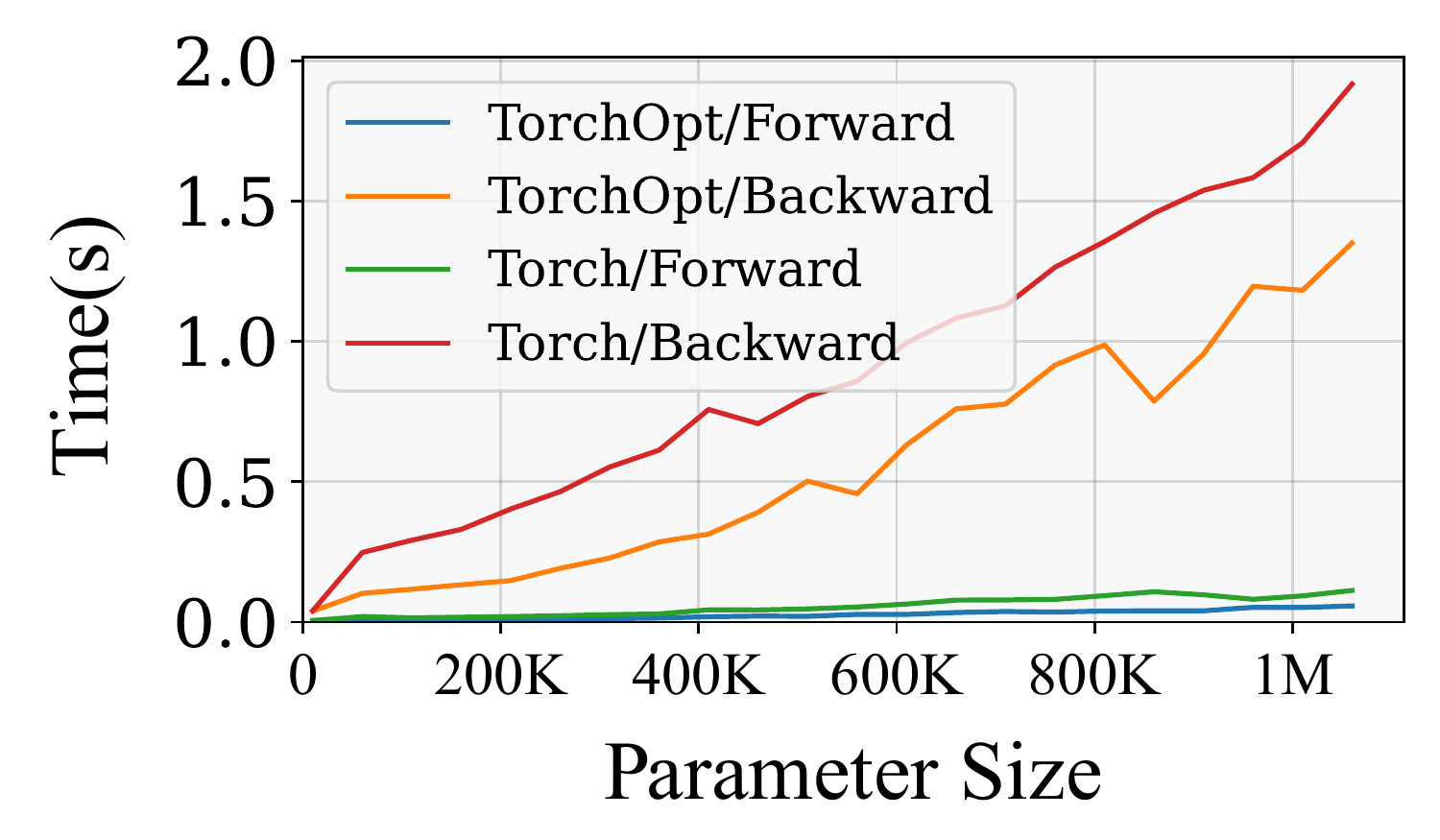}
    \label{fig:perf-overall}}
    \subfigure[GPU-accelerated optimizer]{\includegraphics[width=.31\textwidth]{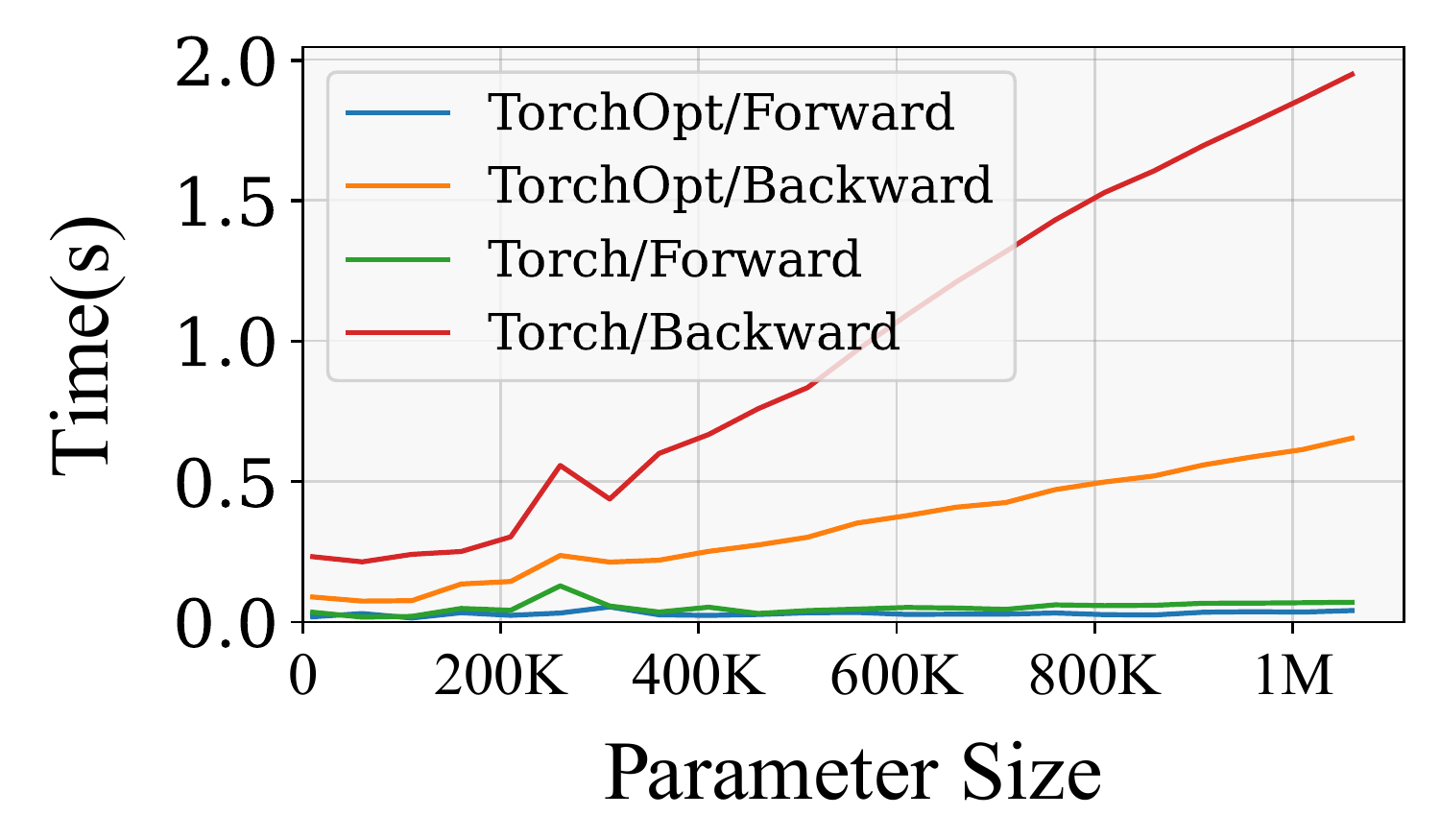}
    \label{fig:perf-meta}}
    \subfigure[Distributed Speedup Ratio]{\includegraphics[width=.31\textwidth]{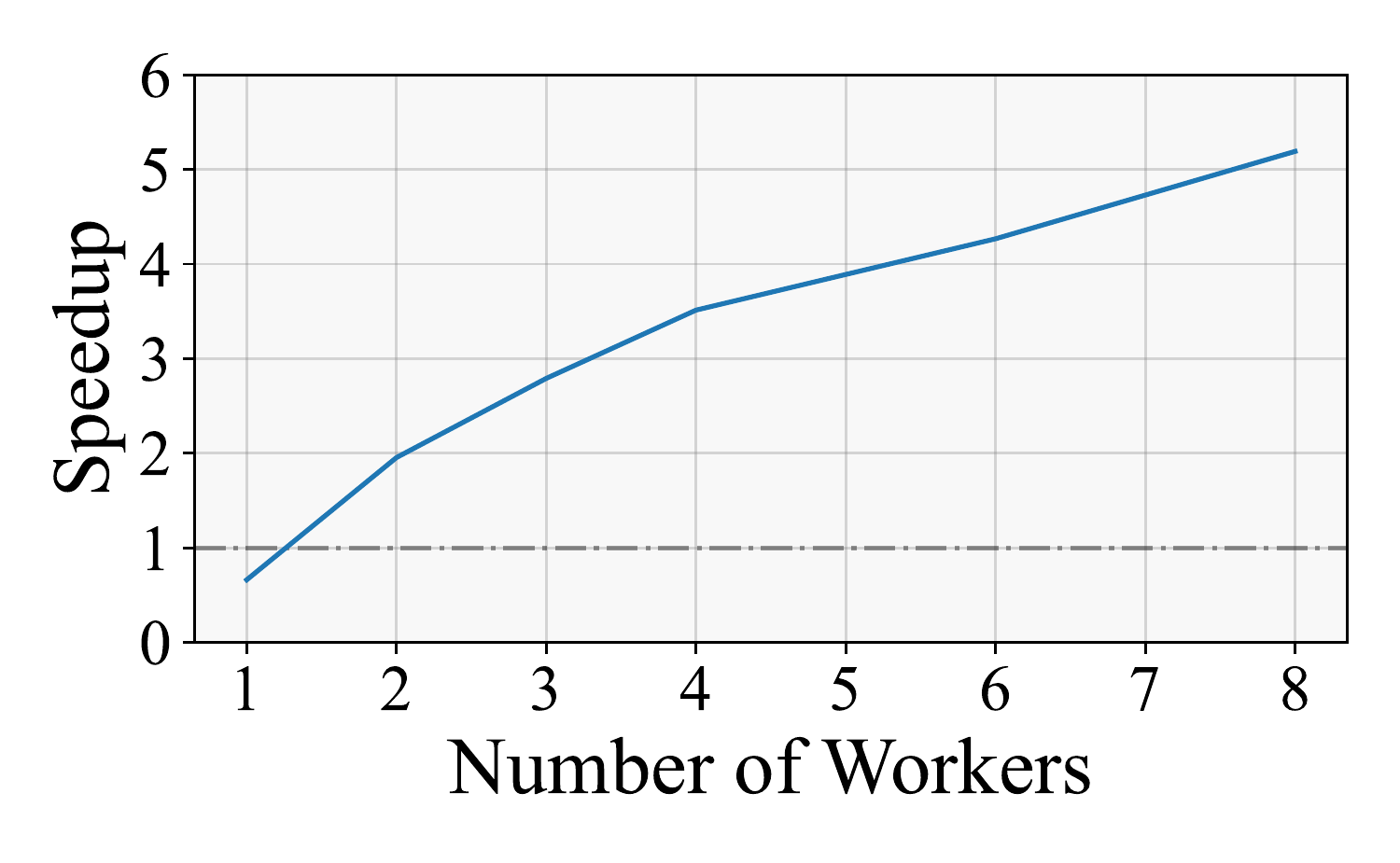}
    \label{fig:distributed}}
    
    \caption{Performance of \torchopt, $(a)$ and $(b)$ are the forward/backward time (Adam optimizer) in different parameter sizes comparing \torchopt and PyTorch, $(c)$ is the speedup ratio on multi-GPUs using RPC compared with the sequential implementation.}
    \vspace{-1.5em}
    \label{fig:results}
\end{figure}

\textbf{CPU/GPU-accelerated optimizers.}
We take the optimizer as a whole instead of separating it into several basic operators (e.g., \texttt{sqrt} and \texttt{div}). Therefore, by manually writing the forward and backward functions, we can perform the symbolic reduction. In addition, we can store some intermediate data that can be reused during the back-propagation. Our design reduces computation and also benefits numerical stability (by explicitly canceling some $0/0$ cases in higher gradient computation). We write the accelerated functions in C++ OpenMP and CUDA, bind them by \texttt{pybind11} to allow Python can call them, and then we define the forward and backward behavior using \texttt{torch.autograd.Function}. The results in Fig.~\ref{fig:results}(a) and Fig.~\ref{fig:results}(b) show that our design largely reduces the optimizer forward and backward time. Refer to Appendix \ref{apx:acc} for experimental results comparing \torchopt and Higher \cite{higher} on the MAML example.

\noindent\textbf{Memory-efficient and cache-friendly PyTree.}
The tree operations (e.g., flatten and unflatten) are frequently called by the functional and Just-In-Time (JIT) components in \torchopt. To enable memory-efficient nested structure flattening, we implement the pytree utilities, named \texttt{OpTree}. By optimizing their memory and cache performance (e.g., \texttt{absl::InlinedVector}), \texttt{TorchOpt} can significantly improve the performance of differentiable optimization at scale. Refer to Appendix \ref{apx:optree} for \texttt{OpTree} experimental results.

\noindent \textbf{Distributed differentiable optimization.}
\texttt{TorchOpt} allows users to reduce training time by using parallel GPUs. Different from existing MPI-based synchronous training~\cite{mai2020kungfu} and asynchronous model averaging~\cite{koliousis12crossbow} systems, \texttt{TorchOpt} adopts RPC as a flexible yet performance communication backend. The distributed GPUs perform differentiable optimization tasks in parallel. These GPUs are coordinated by a chosen GPU device which realizes the synchronous execution of parallel GPUs, thus guaranteeing the convergence of the model in a distributed training setting). 

As shown in Fig.~\ref{fig:rpc}, \texttt{TorchOpt} distributes a differentiable optimization job across multiple GPU workers and executes the workers in parallel. \texttt{TorchOpt} users can wrap code in the distributed Autograd module and achieve substantial speedup in training time with only a few changes in existing training scripts. Fig.~\ref{fig:results}(c) shows that \texttt{TorchOpt} can achieve linear speed-up with MAML when increasing the number of GPU workers (more details in Appendix~\ref{apx:distrain}).
\vspace{-0.1em}
\section{Conclusion and Future work}
This paper introduces \texttt{TorchOpt}, a novel efficient differentiable optimization library for PyTorch. Experimental results show that \texttt{TorchOpt} can act as a user-friendly, high-performance, and scalable library when supporting challenging gradient computation with PyTorch. In the future, we aim to support more complex differentiation modes and cover more non-trivial gradient computation problems, such as adjoint methods for the gradient of ODE solutions, RL or Gumbel-Softmax method for differentiating through discrete distribution, and differentiable combinatorial solvers. \texttt{TorchOpt} has already been used for meta-gradient research problem~\cite{liu2021settling} and we believe it can be served as an important auto-differentiation tool for more differentiable optimization problems.

\acks{We thank Vincent Moens for his constructive comments on \texttt{TorchOpt}.
}

\bibliography{ref}
\newpage
\clearpage
\appendix

\section{Gradient Graph Visualization}

\begin{figure}[ht]
    \centering
    \includegraphics[width=1.1\linewidth]{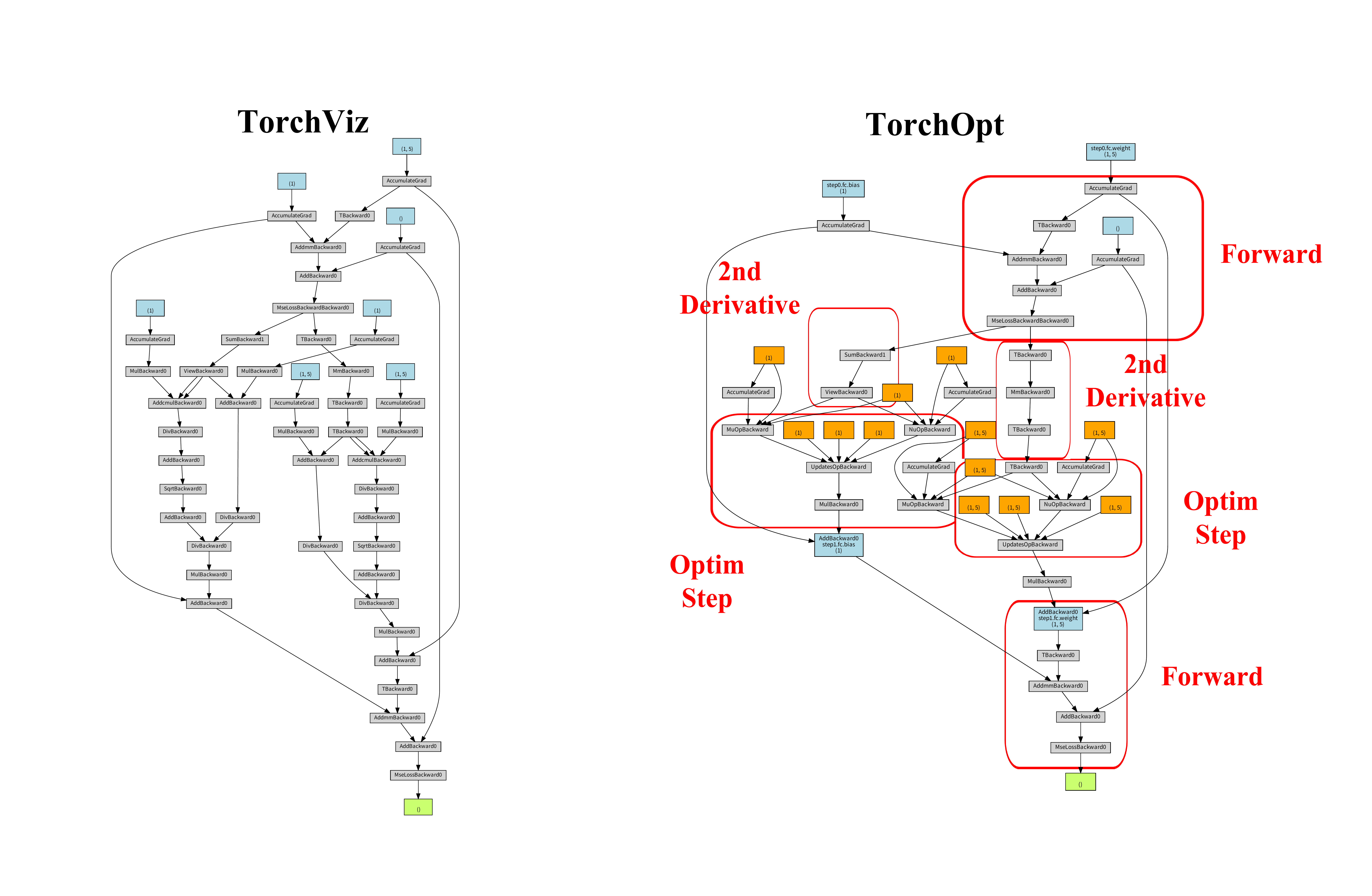}
    \vspace{-1.5em}
    \caption{Gradient graph visualization comparison between TorchViz and TorchOpt.}
    \label{fig:visual}
\end{figure}

Fig. \ref{fig:visual} shows the visualization example of MAML. We use red squares to represent what each part accomplishes separately. Compared with TorchViz, \texttt{TorchOpt} fuses the operations within the Adam together (orange) to reduce the complexity and provides a more straightforward visualization.

\label{apx:visualization}

\section{CPU/GPU-Accelerated Optimizers}
\label{apx:acc}
\begin{figure}[tb]
    \centering
    \vspace{-0.5em}
    \subfigure[CPU-accelerated Meta optimization time]{\includegraphics[width=.49\textwidth]{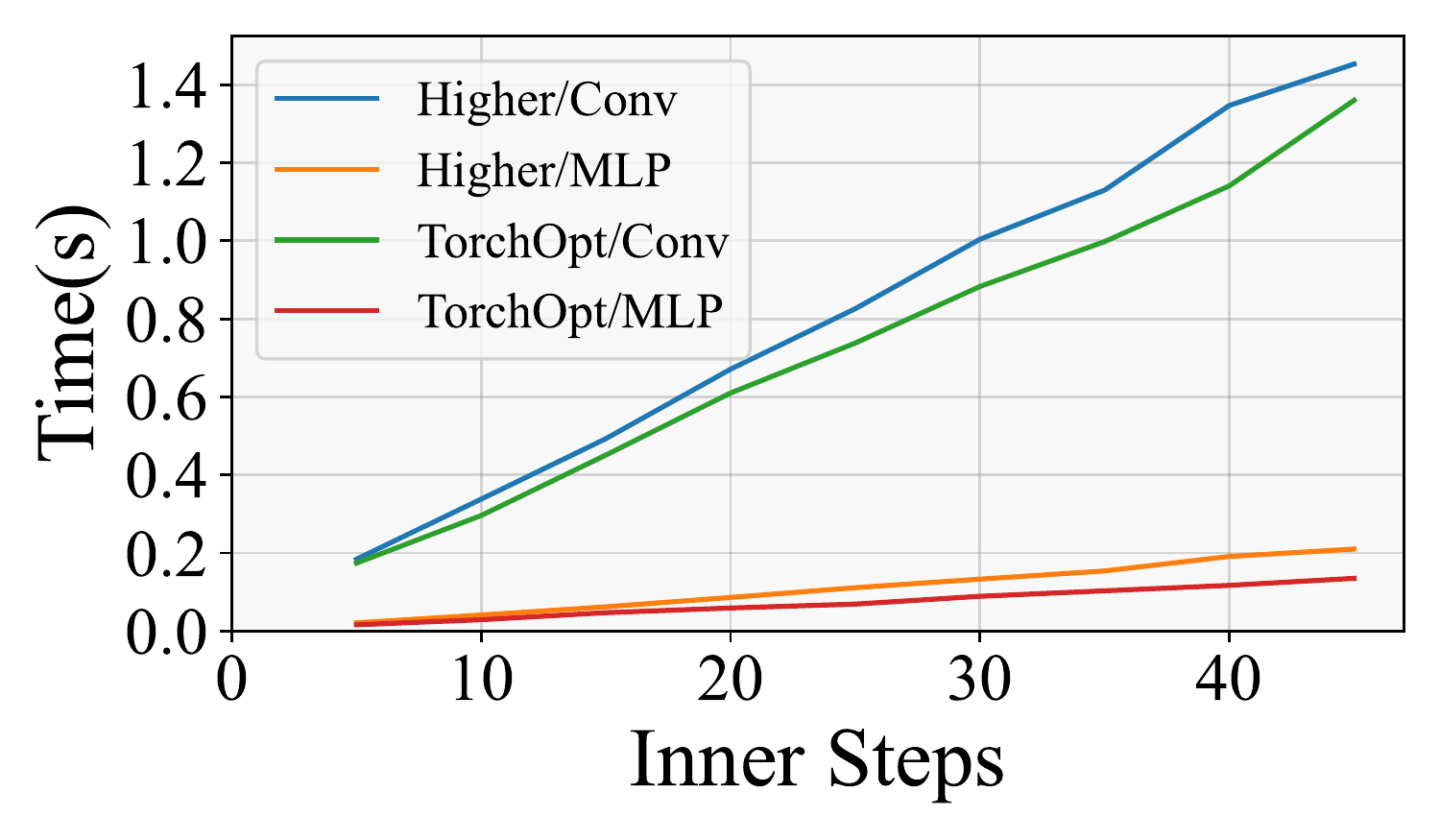}
    \label{fig:meta-optimization-cpu}}
    \subfigure[GPU-accelerated Meta optimization time]{\includegraphics[width=.49\textwidth]{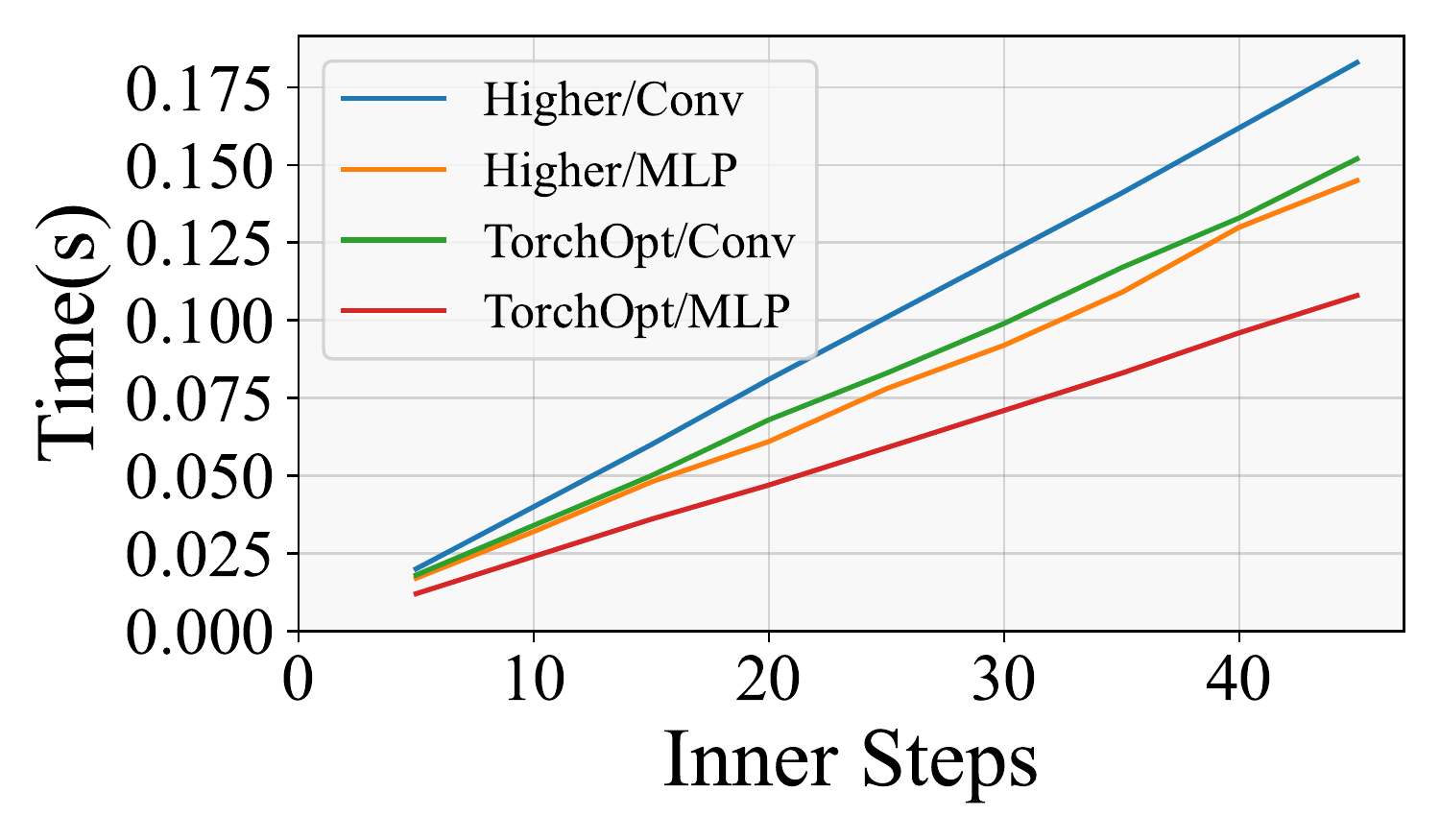}
    \label{fig:meta-optimization-gpu}}
    
    \caption{Performance of \torchopt compared with Higher using MAML example, $(a)$ and $(b)$ are the meta-optimization time (Adam optimizer) in different inner steps and model structures.}
    \vspace{-1.5em}
    \label{fig:meta-optimization}
\end{figure}
Fig. \ref{fig:meta-optimization} shows the meta-optimization time comparison with Higher \cite{higher} in the CPU and GPU settings. Note that the meta-optimization process consists of extra computation beyond the optimizer, where we do not offer acceleration. However, the acceleration is still significant (around $\%25$) for the MLP model in the CPU setting and both Conv/MLP model in the GPU setting.
\section{Distributed Training}
\label{apx:distrain}
\subsection{Distributed Framework}
\begin{figure}[ht]
    \centering
    \includegraphics[width=0.75\textwidth]{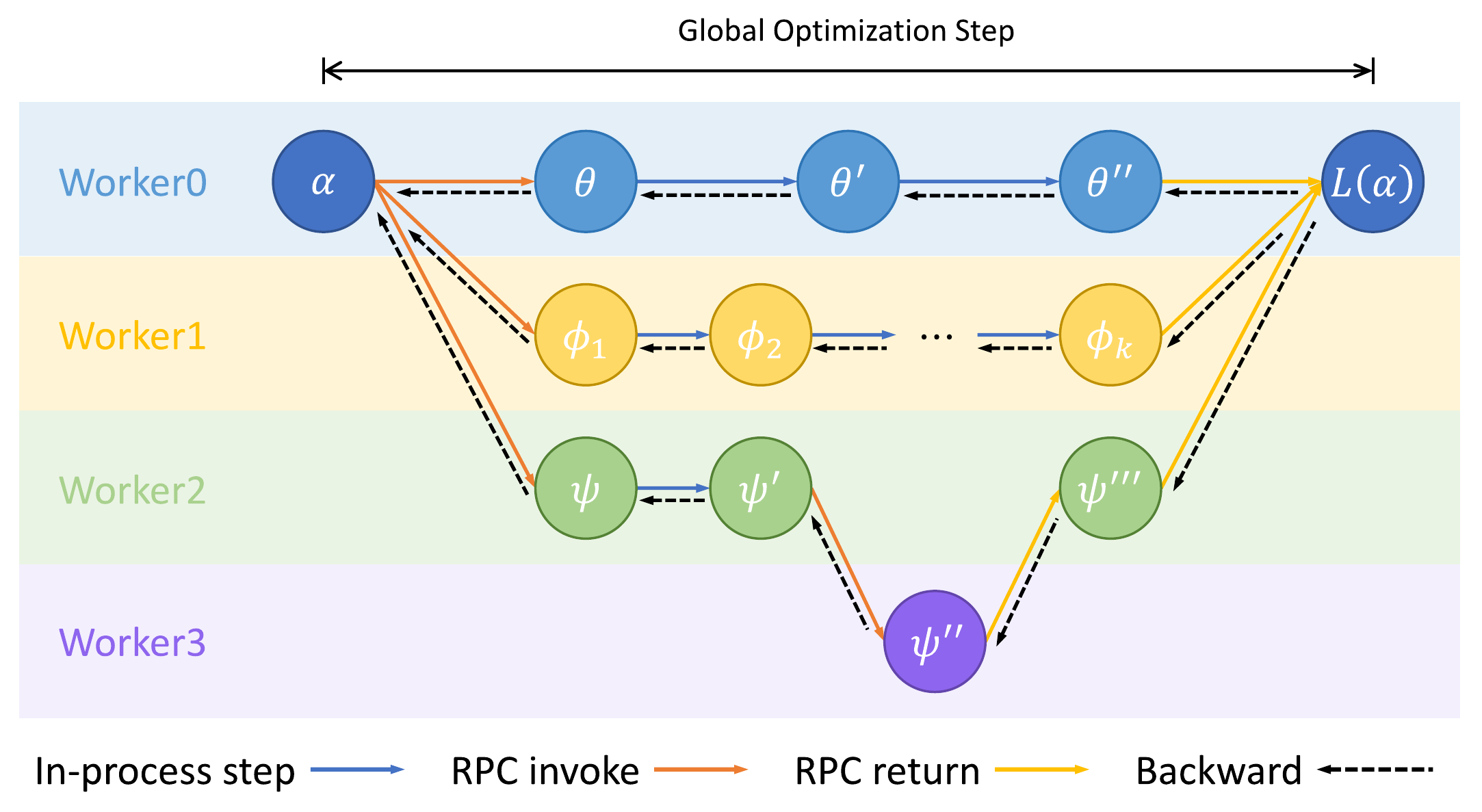}
    \caption{Overview of the Distributed RPC and Autograd framework. The forward and backward pass can be distributed on multiple processes and multiple nodes. The RPC framework supports heterogeneous workloads for different workers.}
    \label{fig:rpc}
    \vspace{-1em}
\end{figure}

In Fig.~\ref{fig:rpc} we show the overview of our distributed framework.
\subsection{Distributed MAML Performance}
\begin{figure}[H]
    \centering
    \includegraphics[width=0.8\linewidth]{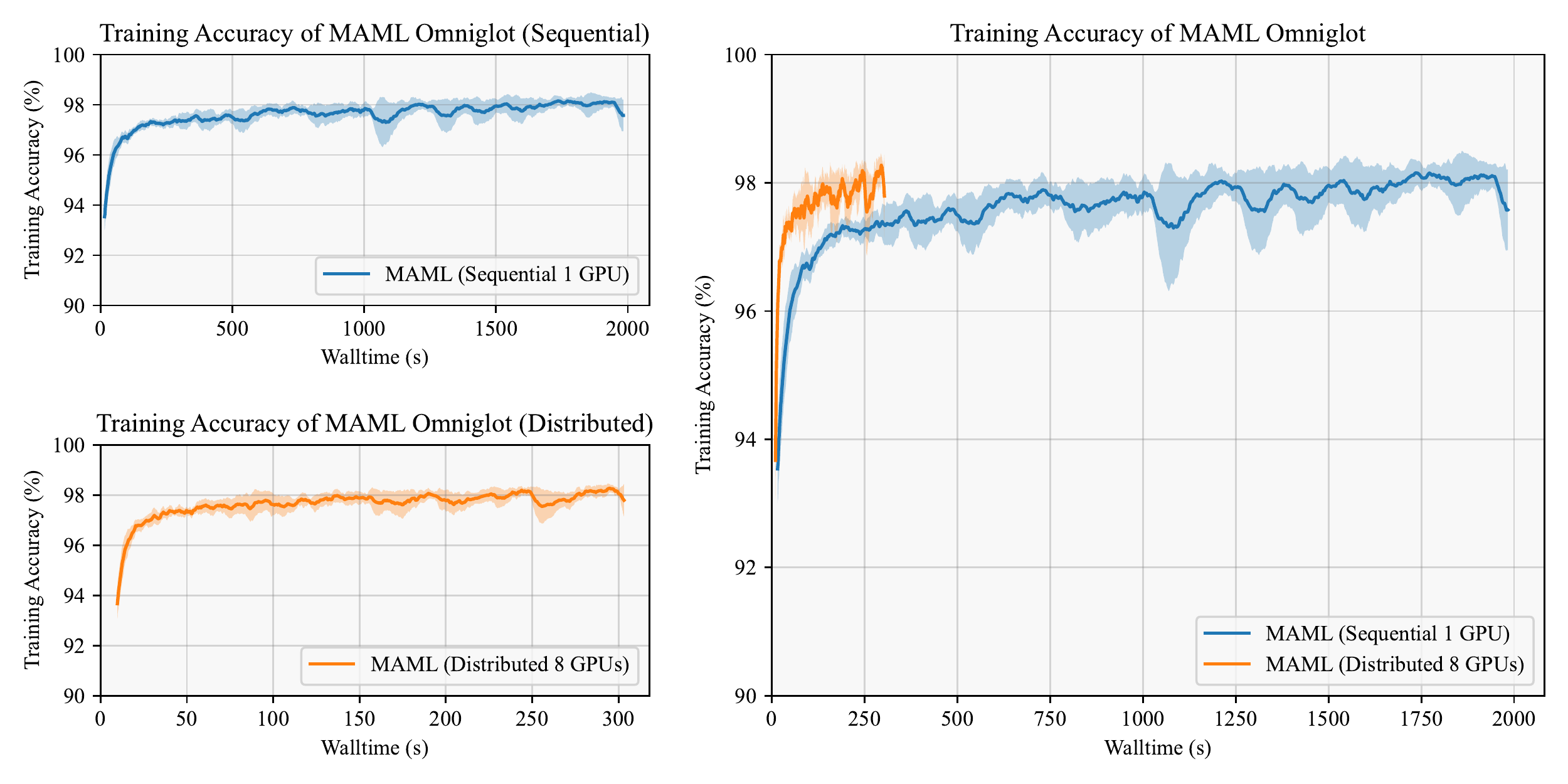}
    \vspace{-1.5em}
    \caption{Wall time comparison between sequential training results and distributed training on 8 GPUs for MAML implemented with TorchOpt.}
    \label{fig:dist-walltime}
\end{figure}

In Fig.~\ref{fig:dist-walltime}, we show the training accuracy and wall time comparison on the MAML Omniglot example. Distributed training achieves better performance and much higher computational efficiency.

\section{OpTree Performance}
\label{apx:optree}

\begin{figure}[H]
    \centering
    \includegraphics[width=0.9\linewidth]{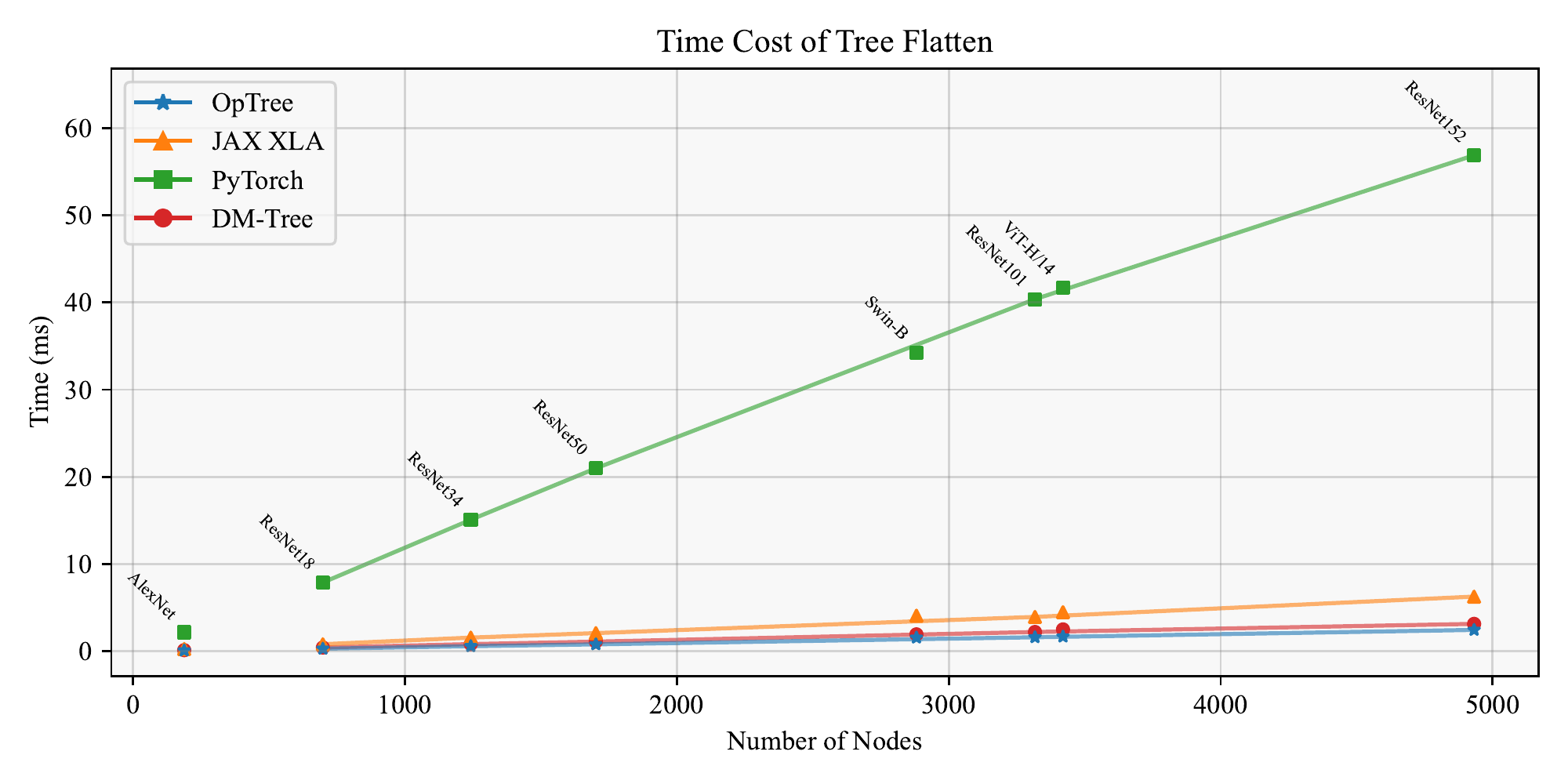}
    \vspace{-1.5em}
    \caption{Tree-Flatten time comparison with respect to the tree scale.}
    \label{fig:tree-flatten}
\end{figure}

\begin{figure}[H]
    \centering
    \includegraphics[width=0.9\linewidth]{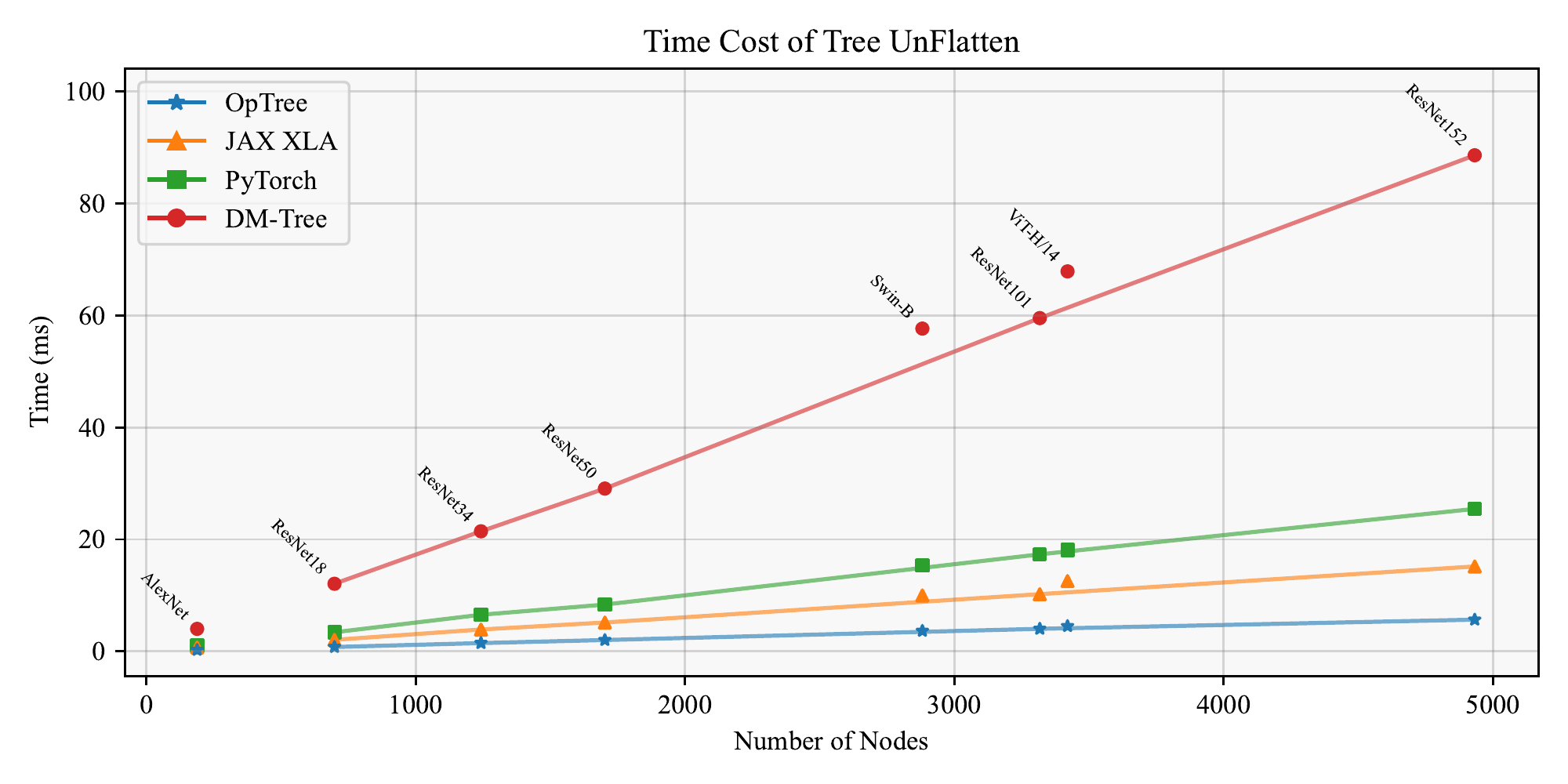}
    \caption{Tree-UnFlatten time comparison with respect to the tree scale.}
    \vspace{-1.5em}
    \label{fig:tree-unflatten}
\end{figure}

\begin{figure}[H]
    \centering
    \includegraphics[width=0.9\linewidth]{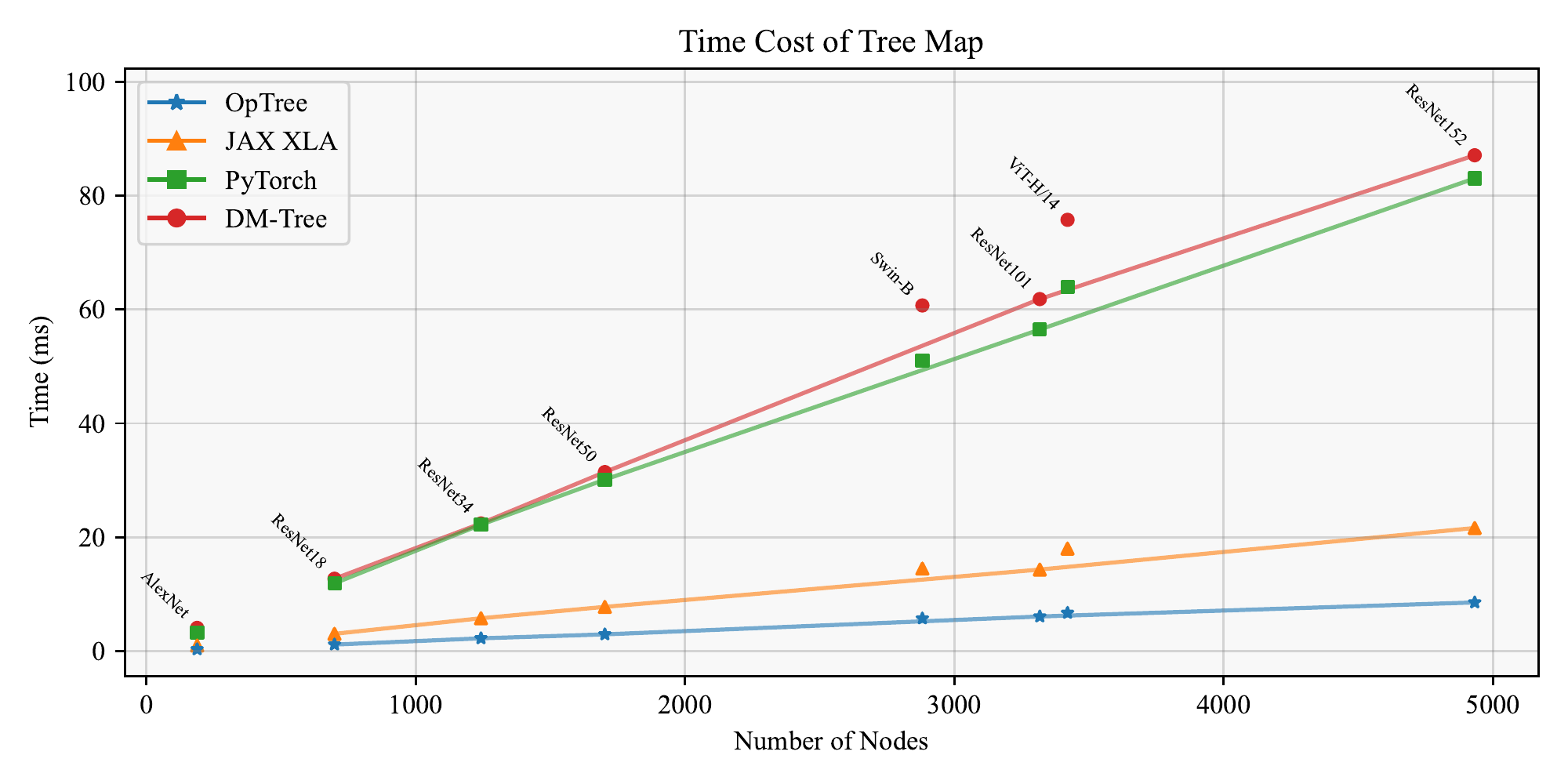}
    \caption{Tree-Map time comparison with respect to the tree scale.}
    \vspace{-1.5em}
    \label{fig:tree-map}
\end{figure}

In Table.~\ref{tab:optree-perf} we show the Speedup ratios of tree operations with ResNet models comparing OpTree, JAX XLA, PyTorch, and DM-Tree. In Fig.~\ref{fig:tree-flatten}, \ref{fig:tree-unflatten} and \ref{fig:tree-map}, we show the time cost of tree-flatten, tree-unflatten, and tree-map trees in a different number of nodes comparing OpTree, JAX XLA, PyTorch, and DM-Tree. OpTree achieves a large speedup compared with all baselines.

\begin{table}[htbp]
    \caption{Speedup ratios of tree operations with ResNet models. Here, \texttt{O}, \texttt{J}, \texttt{P}, \texttt{D} refer to OpTree, JAX XLA, PyTorch, and DM-Tree, respectively.}\label{tab:optree-perf}
    \centering
    \footnotesize
    \begin{tabular}{c|ccc|ccc|ccc|ccc}
        \toprule
        Module Scale   & \multicolumn{3}{c|}{ResNet18} & \multicolumn{3}{c|}{ResNet50} & \multicolumn{3}{c|}{ResNet101} & \multicolumn{3}{c}{ResNet152}                                                                                                                                                                                                                 \\
        Speedup Ratio  & \texttt{J} / \texttt{O}       & \texttt{P} / \texttt{O}       & \texttt{D} / \texttt{O}        & \texttt{J} / \texttt{O}       & \texttt{P} / \texttt{O} & \texttt{D} / \texttt{O} & \texttt{J} / \texttt{O} & \texttt{P} / \texttt{O} & \texttt{D} / \texttt{O} & \texttt{J} / \texttt{O} & \texttt{P} / \texttt{O} & \texttt{D} / \texttt{O} \\
        \midrule
        Tree Flatten   & 2.80                          & 27.31                         & 1.49                           & 2.63                          & 26.52                   & 1.40                    & 2.46                    & 25.18                   & 1.38                    & 2.56                    & 23.25                   & 1.28                    \\
        Tree UnFlatten & 2.68                          & 4.47                          & 15.89                          & 2.56                          & 4.16                    & 14.51                   & 2.55                    & 4.32                    & 14.86                   & 2.68                    & 4.51                    & 15.70                   \\
        Tree Map       & 2.61                          & 10.17                         & 10.86                          & 2.63                          & 10.18                   & 10.62                   & 2.35                    & 9.26                    & 10.13                   & 2.53                    & 9.69                    & 10.16                   \\
        \bottomrule
    \end{tabular}
\end{table}

\vspace{-1em}
    
\vspace{-1em}

\end{document}